# Investigating the Effect of Technostress on the Perceived Organizational Commitment by Mediating Role of Individual Innovation.

**Hassan Hessari\***, Department of Business Information, Technology, Pamplin College of Business, Virginia Tech, Blacksburg, Virginia, USA hassanhessari@vt.edu

**Fatemeh Daneshmandi,** Academic Center for Education, Culture, and Research, Neyshaboor, Iran s.daneshmandi69@gmail.com

**Tahmineh Nategh**, Department of Management, Shahrood Branch, Islamic Azad University, Shahrood, Iran tnategh@iau-shahrood.ac.ir

**Abstract:**

**Purpose:** Technology plays a pivotal role in shaping the fate of organizations, both positively and negatively. One of its detrimental consequences is the emergence of "Technostress," a form of destructive stress. This paper investigates the impact of technostress on Perceived Organizational Commitment (POC) through the lens of individual innovation. The objective is to provide valuable insights for organizational managers, enabling them to effectively mitigate the adverse effects of technostress within their teams.

**Design/Methodology/Approach:** This study utilized a questionnaire survey conducted within an Engineering Consulting Company in Iran, with 147 individuals participating, selected according to Morgan's table.

**Findings:** The research findings revealed three crucial insights: (1) Technostress significantly and negatively influences both POC and individual innovation. (2) Individual innovation positively and significantly impacts POC. (3) Individual innovation acts as a mediator between technostress and POC, alleviating the negative impact of technostress on organizational commitment.

**Research Implications:** The study underscores the importance for managers to proactively address technostress-related challenges and promote individual innovation within their organizations. These efforts are vital in enhancing organizational commitment among employees.

**Originality/Value:** This research makes a significant contribution to the field by illuminating the mediating role of individual innovation in the relationship between technostress and perceived organizational commitment. Given the close association of employees in engineering organizations with technology, this study sheds light on the specific challenges faced by this sector, thereby enhancing our understanding of technostress effects in the workplace.

## 1. Introduction

The rapid proliferation of information and communication technologies in organizations has brought about significant improvements in employee performance, job satisfaction, productivity, and overall effectiveness (DeLone & McLean, 2003; Hessari & Nategh, 2022a; Mohammadi et al., 2023; Sasani et al., 2023). However, alongside these positive effects, technology also has its drawbacks (Hessari & Nategh, 2022b). One of the most pressing issues faced by organizations today is technostress, a form of stress associated with the use of information technology (IT), or the demands placed on individuals by IT (Maier et al., 2015). This psychological state of stress is often accompanied by physical and biological symptoms, such as increased arousal in employees who extensively use computers for work (Arnetz & Wiholm, 1997; Riedl, 2012). Technostress has emerged as a crucial research topic due to its adverse consequences on employees, their families, and the organizations they work for (D'Arcy et al., 2014; Tarafdar et al., 2013).

One of the significant and contentious impacts of technostress is its effect on employees' organizational commitment. Understanding the value and importance of organizational commitment is paramount; committed employees are those deeply involved in the organization, passionately dedicated to staying with it. Such employees are less likely to resign or be absent and are more inclined to share knowledge and contribute to the organization's success (Pratama et al., 2022). Therefore, organizational commitment plays a pivotal role in the overall functioning of any organization.

Innovation, a crucial factor in today's fast-paced world, has garnered substantial attention from researchers examining its influence on organizational commitment. Innovation is intricately linked with flexibility and productivity and holds immense significance for employees, organizations, and societies at large (Saunila, 2020). The ability to adapt to innovative technologies and methods is essential for employees, as it directly impacts an organization's efficiency and competitiveness in the current sensitive and competitive landscape.

Technostress affects various industries, with the engineering sector being one of the most consistently impacted (Hessari & Nategh, 2022c). Engineering organizations heavily rely on sophisticated software, which is continuously updated to align with the progress in scientific and engineering principles. The rules and principles governing these software applications undergo frequent changes, necessitating employees to constantly update their skills. This adaptability is vital for enhancing an organization's efficiency and performance in today's highly competitive environment.

## 2. Theoretical foundation
### 2.1. Technostress

Technostress constitutes a significant challenge for both organizations and their employees. It represents a type of stress that individuals experience due to their reliance on technology, coupled with the anxiety arising from uncertainties related to technology usage. The term was coined by Craig Brod, a consultant and psychologist specializing in adapting to new technology, in his seminal work "Technostress: The Human Cost of The Computer Revolution". Brod conceptualized technostress as a modern ailment resulting from the human struggle to adapt healthily to rapidly evolving global computer technologies. While Brod viewed technostress as an illness, other scholars argue that it signifies an inability to adapt to the changes ushered in by technology. Consequently, this inability affects individual efficiency and, consequently, organizational success (Dragano & Lunau, 2020).

In today's fast-paced technological landscape, particularly in the realm of information technology, rapid changes render outdated organizational practices obsolete, making it challenging for such organizations to compete effectively. Embracing new technologies, although essential, can induce stress among employees, potentially hampering their efficiency. Despite the costs associated with implementing Information Technology (IT) solutions, they are indispensable for organizations. IT enables timely access to new information, facilitating well-informed and appropriate decision-making (Tarafdar et al., 2007; Weil & Rosen, 1997).

Drawing from data in the United States, Tarafdar and colleagues expanded the technostress scale measurement. They delineated technostress into five components, which are crucial in understanding the various dimensions of this phenomenon (Tarafdar et al., 2007). These components shed light on the multifaceted nature of technostress, providing valuable insights for organizations aiming to mitigate its adverse effects on both individuals and the overall organizational performance (See Table 1).

Table (1): Technostress Component (Tarafdar et al., 2007).

| Technostress Component | Description | Summary |
|---|---|---|
| Techno-Overload | When an individual is forced to work at higher speeds and for longer hours because of technology | Too Much |
| Techno-Invasion | When an Individual can be contacted anytime and the boundaries between personal and work hours become less distinct | Always Connected |
| Techno-Complexity | When an individual feels that their skills are not sufficient due to difficult ICT | Difficult |
| Techno-Insecurity | When an individual feels their job is under threat by ICT technologies or by someone who is better using ICT | Uncomfortable |

| Techno-Uncertainty | When an individual is hesitant/disturbed as technology is always changing/upgrading | Too often and Unfamiliar |

## 2.2. Perceived Organizational Commitment

In the contemporary business landscape, organizations have recognized the pivotal role played by human resources as the primary source of gaining competitive advantages (Rouhani & Mohammadi, 2022). The current competitive arena sees organizations vying for supremacy through the provision of comprehensive welfare programs and the meticulous attention dedicated to staff recruitment and retention (Nawaser et al., 2014). This trend is highlighted in Fortune magazine's recent ranking of 100 companies, underscoring the critical importance of fostering a favorable working environment for human resources (Riggle et al., 2009).

The concept of commitment has been a longstanding focus in management literature, leading to diverse definitions, models, and measurement methods over the years (Nawaser et al., 2011). Organizational commitment (OC) has garnered significant attention from scholars and practitioners alike, shaping various behaviors and attitudes within the organizational context (Chen & Francesco, 2003; Cheng & Stockdale, 2003; Meyer et al., 2002). Defined as a strong desire for an individual's survival within a specific organization, the willingness to invest substantial effort for the organization, and the unequivocal acceptance of the organization's values and goals (Clugston et al., 2000), OC has undergone substantial evolution.

The evolution of OC can be segmented into distinct periods, each profoundly influencing its conceptualization and applications (Cohen, 2007; Meyer & Herscovich, 2001; Nawaser et al., 2015). Initially, research focused on Howard Becker's concept of side-bets, representing valuable investments individuals accrue and stand to lose upon departure (Cohen, 2007). Subsequently, emphasis shifted from material side benefits to psychological attachment, particularly in the work of Allen and Meyer (1990), who conceptualized OC as a unidimensional construct solely centered on affective attachment. In the last period, pioneering work by Allen and Meyer (1991) led to the development of vital multi-dimensional approaches. Their three-element model of organizational commitment encompassed continuous, normative, and affective commitment as components of Attitude Commitment (Powell & Myer, 2004). While initial OC definitions might have appeared conceptual and descriptive, theorists have operationalized it as a multidimensional phenomenon, marking a significant step in understanding organizational commitment. In the subsequent sections, a comprehensive exploration of OC dimensions will be undertaken, particularly from the viewpoint of Allen and Meyer's influential contributions.

According to Allen and Meyer (1990), affective commitment is the emotional attachment of an individual to the organization and their identification with it. It consists of three aspects: affective affiliation to the organization, individual's identification through the organization, and tendency to continue working in the organization. Continuous commitment is another dimension of organizational commitment. It is based on Becker's investment theory, which states that over time, an individual accumulates a capital in the organization that increases with their experience and that losing it would be costly. Meyer et al. (1989) argue that continuous commitment is the individual's psychological attachment to the organization resulting from the employee's perception of the costs of leaving the organization. In other words, continuous commitment stems from the individual's awareness of the consequences of quitting the organization. Lastly, normative commitment is a sense of obligation to continue working with the organization. Those who have a high level of normative commitment feel that they must stay in the organization (Meyer & Allen, 1997).

Organizational Commitment has a potentially serious impact on the performance of the organization and can predict its effectiveness, so ignoring it might be costly and impose high expenses (Benkhoff, 1997). The effects of OC are identified in two dimensions: individual and organizational. One of its main impacts is the rate of employee's absence from work. Allen and Meyer (1990) suggest that the intersection of the three dimensions of OC is the link between the individual and the organization and consequently reduces turnover. The negative relationship between turnover and OC is also confirmed by Al-Jabari and Ghazzawi (2019). Moreover, the influence of OC has been consistent in different parts of the world; for example, Siu (2003) has verified that OC has a positive relationship with job performance among Hong Kong employees. When an individual is committed to the organization, their rate of absence or turnover will decrease, which will lead to lower costs and higher efficiency for the organization (Meyer et al., 2002). Furthermore, the individual dimension of OC has various impacts such as: friendship, assistance to colleagues, and reduction of job stress (Wasti, 2005). Also, those who showed more commitment to the organization were more loyal, had less job stress (Muthuveloo & Rose, 2007), and were more likely to accept organizational changes (Vakola & Nikolaou, 2005). The positive outcomes of OC are briefly presented in Table 2.

Table (2): OC's results (Laka-Mathebula, 2005).

| Levels of Analysis | | |
|---|---|---|
| **Individual** | **Groups** | **Organization** |
| Feeling of attachment and affiliation | Stability and membership | Effectiveness increase |
| Feeling of security | Consistency and solidarity | Reduction of job leaving and displacement |
| Positive impression of yourself and feeling proud | Group effectiveness | Reduction of delays and absence |
| Individual efforts and attempts | Conflict reduction | Attractions for outsiders Efficiency increase |

## 2.3. Individual Innovation

Chen et al. (2004) define innovation as the introduction of new combinations of essential factors in production in a manufacturing system. Innovation capital is "organizational competence, conducting research and development, and creating new technology and product in order to meet customer's demand". The Organization for Economic Cooperation and Development (OECD) considers any kind of commercial exploitation of new knowledge as innovation. In other words, it is related to the implementation of new processes or procedures that are significantly different from the existing ones (Du Plessis, 2007).

Innovation includes six different operations: new product, new service, new ways of production, finding new markets, new supply sources, and new ways of organizing (Johannessen et al., 2001). Recently, it has been widely accepted among researchers and scholars that innovation brings power for organizations (Drach-Zahovy et al., 2004). Economists, researchers, and managers believe that innovation is the criterion for the differentiation of societies and the creation of competition among them, and that it is the prerequisite for prosperity and stability in every society. Fumio Kodama, a Japanese researcher, states that "increasing productivity and speed of innovation in society is the determinant factor of economic health in all societies" (Miller & Morris, 2008). To effectively use the resources, increase productivity, expand global trade, improve individual and social welfare, or in other words increase the level of living standards, investigation on innovation is highly needed (Miller & Morris, 2008). Undoubtedly, individual innovation of employees is an important issue in every organization. In recent years, the key to success and survival of any organization is employee's innovation and encouraging him/her to innovate. Individual innovation is at the core of many principles in successful management such as total quality management, continuous improvement plan, corporate's boldness, creative problem solving, organizational learning (Osayawe Ehigie & Clement Akpan, 2004).

Employees are considered as vital factors in innovation realization. The word "innovation" is derived from the word "invention" which is rather rare, while incremental innovations based on employee effort are more common. Individual innovation is, in fact, at the heart of many principles in successful management such as total quality management (Osayawe Ehigie & Clement Akpan, 2004), continuous improvement plan (Fuller et al., 2006), corporate boldness (Elfring, 2005), creative problem solving (Basadur, 2004), organizational learning (Senge, 2006). Previous studies indicate that individual innovation is beneficial for any organization, a positive relationship was empirically observed between specific innovation behaviors and organizational performance (Miron et al., 2004). It is also revealed that individual innovation does not diminish the quality and efficiency of routines. Employees are fully capable of maintaining the balance between innovation and required organizational features in promoting innovation. Quality and productivity are complementary not competitors. Getz and Robinson (2003) proposed an interesting principle from rule of thumb: corporations that use 80% of employee's ideas and 20% of planned innovation activities would be more advanced. It is confirmed that organization's competence in achieving new products and other aspects of performance is related to human resources' knowledge of that organization (Foss, 2007). Therefore, the most unique and irreplaceable resources available for companies are those individuals who have the knowledge and the potential of effectively using other resources of the organization (Argote & Ingram, 2000).

### 3. Hypotheses development

According to Moore (2000), Sethi et al. (1999), and Murray and Rostis (2007), technology may frustrate or bore the employees, which leads to a reduction of efficiency. Managing technostress can be challenging for an organization. Weil and Rosen (1997) found that technostress causes excessive job demands, information overload, loss of motivation, and job dissatisfaction. Furthermore, Marchiori et al. (2020) investigated this issue in their paper titled "A relationship between technostress, satisfaction at work, organizational commitment and demography: Evidence from the Brazilian public sector." The results showed that perceived job stress may decrease job satisfaction, while technology management strategy leads to more job satisfaction and OC. Moreover, Kumar et al. (2013) also examined this concern in their paper titled "Impact of technostress on job satisfaction and organizational commitment among IT professionals". They selected 80 professionals from a Technology Park. The obtained results indicated that there was a negative correlation between OC and job satisfaction. Kumar et al. (2013) proposed the following hypothesis:

- H1: Technostress has negative and meaningful impact on OC.

In their study employing structural equation modeling, Mani et al. (2014) explored the intricate interconnections among organizational environment, work-family conflict, job stress, individual characteristics, job satisfaction, and organizational commitment (OC) within the context of US pharmacists. The research highlighted the pivotal role of specific variables in influencing job stress, with excessive workload, role ambiguity, and role conflict emerging as the most significant stressors. Additionally, the study revealed a noteworthy correlation between individual characteristics, job satisfaction, and OC (Mani et al., 2014). Based on the study's findings, it can be deduced that individual characteristics exert a substantial and adverse impact on job stress. Given the analogous nature of technostress and job stress, a hypothesis is posited:

- H2: Technostress has negative and meaningful impact on individual innovation.

In their comprehensive study titled "Organizational and Professional Commitment among Scholars and Engineers: Innovation Management," Perry et al. (2016) delved into the intricate relationship between innovation, innovation management, and organizational commitment (OC). Conducted among a cohort of 225 engineering researchers, the research shed light on a pivotal nexus between OC and innovation within organizational settings. The study's meticulous analysis underscored a compelling and positive correlation between employees' proclivity for innovation and their level of OC, indicating that a strong sense of commitment significantly enhances an individual's willingness to innovate and contribute positively to the organization (Perry et al., 2016).

Organizational commitment transcends the boundaries of mere presence within the workplace; it permeates various facets of organizational dynamics (Mohaghar et al., 2022). Committed individuals, particularly those with affective commitment, not only fulfill their job roles but also proactively engage in activities that align with the organization's goals and values. Their dedication goes beyond the norm, reflecting in enhanced job performance and increased organizational productivity. Recent empirical studies echo these observations, showcasing the tangible benefits of a committed workforce. Research conducted by Mowday et al. (2013) as well as Boshoff and Mels (1995) illuminates the positive impact of OC on job performance and employee productivity, substantiating the vital role that commitment plays in organizational success.

Building on this foundation, Meyer et al. (2002) formulated a hypothesis that further strengthens the understanding of OC's influence on organizational dynamics. Their hypothesis posits that individuals deeply committed to their organizations not only exhibit enhanced job performance but also act as catalysts for fostering a positive work environment, thereby contributing significantly to the organization's overall effectiveness and success. This insight underscores the pivotal role of organizational commitment in shaping not only individual behaviors but also the collective ethos of the workplace, emphasizing its importance as a driving force behind innovation, productivity, and overall organizational performance. Therefore, we consider following hypotheses:

- H3: Individual innovation has positive and meaningful impact on OC.
- H4: Individual innovation plays a mediating role between technostress and OC.

Considering theoretical-foundation review and research background, the following model is presented (Figure 1).

Figure (1): Conceptual Model

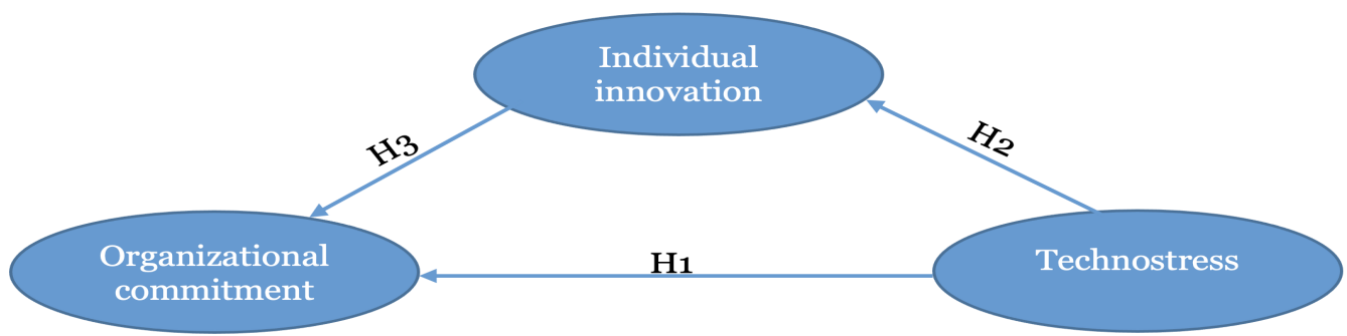

## 4. Methodology

The methodology employed in this study can be characterized as applied, with a focus on understanding specific real-world phenomena. The data collection methodology was descriptive-correlative, implemented using structural equations modeling. Data were gathered using standardized questionnaires distributed in person.

### 4.1. Study Context

This research was conducted within an engineering consulting organization. This choice was made due to the organization's nature of work, which involves diverse activities closely tied to technology. Given the prevalence of technology in engineering consulting organizations, this context provided a rich environment for the study. The research population comprised 240 employees. Using the Morgan Table (Krejcie & Morgan, 1970), a random selection of 170 questionnaires was distributed among the employees. After data collection, 147 questionnaires were found suitable and utilized for analysis. The participants were chosen based on the guidelines provided by the Morgan Table for the second half of the year 2020. Table 3 presents a breakdown of the sample in terms of gender, university degree, and work experience, showcasing the diversity of the participants in the study.

Table (3): Table of frequency and its percentage for research samples.

| Variable | Sub-group | Frequency | Frequency percentage |
|---|---|---|---|
| Sex | Male | 95 | 64/6 |
| | Female | 52 | 35/4 |
| Age | 22 to 30 | 41 | 27/9 |
| | 31 to 39 | 66 | 44/9 |
| | 40 to 48 | 28 | 19/0 |
| | 49 to 57 | 12 | 8/2 |
| Educational level | High school | 11 | 7/5 |
| | Bachelors | 62 | 42/2 |
| | Masters | 66 | 44/9 |
| | PhD | 8 | 5/4 |
| Work experience | 1 to 4 | 25 | 17/0 |
| | 5 to 9 | 46 | 31/3 |
| | 10 to 14 | 37 | 25/2 |
| | 15 to 19 | 22 | 14/9 |
| | 20 to 40 | 17 | 11/6 |

### 4.2. Measures

In this study, various tools were employed to assess the constructs under investigation. The Organizational Commitment (OC) Questionnaire developed by Allen and Meyer (1990) was utilized, comprising eight questions each for affective commitment, normative commitment, and continuous commitment. Additionally, the Technostress Questionnaire by Tarafdar et al. (2007) was administered, encompassing six questions on overload resulting from technology, three questions on stress from technology, five questions on technology's complexity, five questions on insecurity arising from technology, and four questions on uncertainty in technology. Furthermore, the Individual Innovation Questionnaire developed by Janssen (2004) was employed, consisting of three questions each for idea production, idea development, and idea realization. Participants responded on a Likert's five-point scale, ranging from "completely disagree" to "completely agree," with scores ranging from 1 to 5. Following data collection, the obtained data from the questionnaires were processed using SPSS19 and SMART-PLS 2 software for analysis. Descriptive statistics such as frequency and percentage were employed, and inferential statistics, including partial least squares structural equation modeling (PLS-SEM), were conducted using SMART-PLS software.

## 5. Data Analysis and Results

To ensure the validity of the instruments, expert opinions from professors and academic experts were sought. The questionnaire items were designed based on established standards, ensuring their reliability. To assess the reliability of the scales, Cronbach's coefficient alpha was calculated, as shown in Table 4. These reliability measures underscore the robustness and consistency of the instruments employed in this research. Using data obtained from this questionnaire and SPSS software, trust coefficient was measured through Cronbach's Alpha which was estimated more than 0.7 in this paper.

Table (4): The results of the reliability of the questionnaire.

| Questionnaire | Cronbach's coefficient alpha |
|---|---|
| Organizational commitment | 0.970 |
| Technostress | 0.962 |
| Individual innovation | 0.962 |

In this study, the data analysis was conducted using partial least squares structural equation modeling (PLS-SEM), a statistical method employed to explore linear relationships between latent variables and observed variables. Partial least squares structural equation modeling (PLS-SEM) integrates the Confirmatory Factor Analysis (measurement model) and Regression or Path Analysis (structural model), offering a comprehensive approach for simultaneous hypothesis testing. This powerful technique enables researchers to validate or adapt hypothetical structures using empirical data. For our analysis, SMART-PLS software was utilized, as it accommodates complex structural equation models with multiple variables, encompassing direct, indirect, and interactive effects. The analysis involved examining the outputs presented in Table 5, detailing the investigation of research hypotheses. These findings were instrumental in determining the positive or negative sign of the T values derived from the Structural Equation Modeling within the standard estimation model, as depicted in Figures 2 and 3. The results provide valuable insights into the relationships between variables, offering a robust foundation for the conclusions drawn in this study.

Figure (2): Structural equation model in standard estimation model.

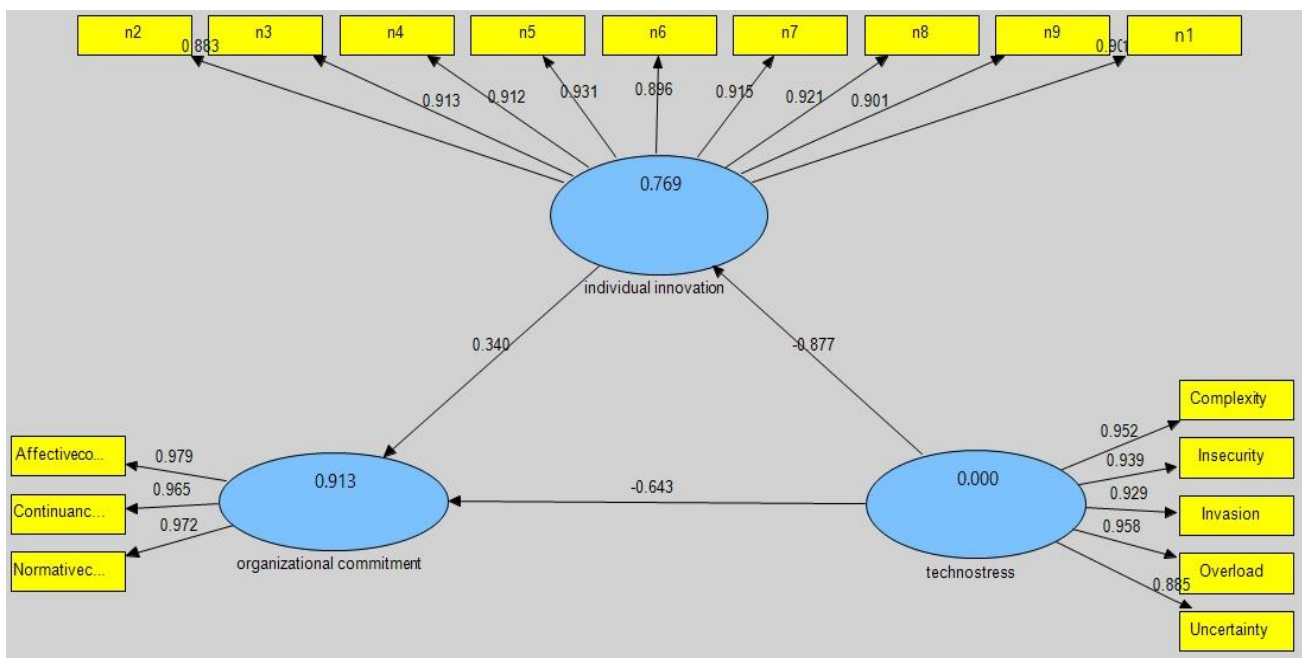

Figure (3): Structural equation model in the case of meaningful coefficients.

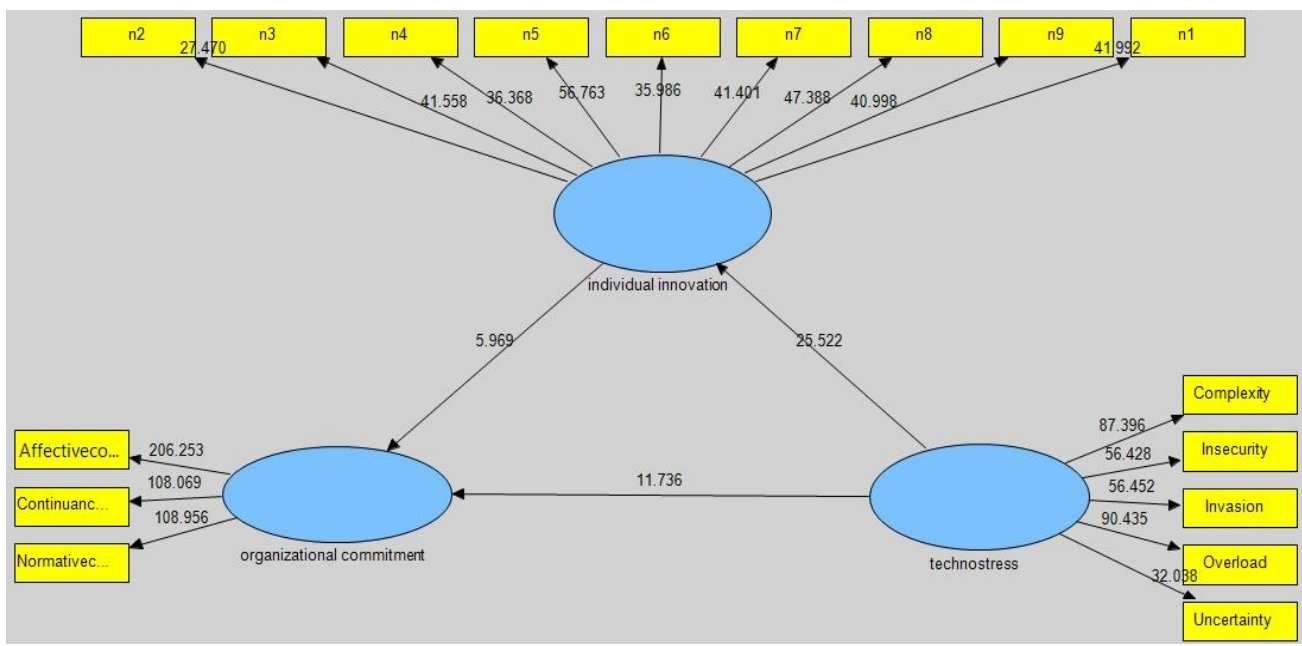

Table (5): test results regardless of impact of the mediating variable.

| Direction | t-Value | Standard | Meaningful level | impact rate |
|---|---|---|---|---|
| Technostress -> OC | -11.736 | -0.643 | Meaningful | Negative |
| Technostress -> Individual innovation | -25.522 | -0.877 | Meaningful | Negative |
| Individual innovation -> OC | 5/969 | 0.34 | Meaningful | Positive |

### 5.1. Evaluating the Mediating Effect

This hypothesis centers on exploring the mediating role of individual innovation in the relationship between technostress and perceived Organizational Commitment (OC). To assess the impact of individual innovation, Baron, and Kenny's test (1986) was employed, employing SPSS software and the Regression method. As indicated in Table 6, significant correlations were found between technostress and perceived OC, technostress, and individual innovation, as well as individual innovation and perceived OC (initial three conditions). In the fourth step, upon introducing the interface variable, the standard beta reduced from 0.940 to 0.644 in the relationship between technostress and perceived OC yet remained statistically significant. This reduction indicates a modest mediating effect of individual innovation. Thus, the fourth hypothesis is substantiated. Detailed results from the hypothesis investigation are presented in Table 7, providing a comprehensive overview of the mediating role of individual innovation in the context of technostress and perceived OC. To gauge the indirect effect of technostress on perceived Organizational Commitment (OC), the path coefficient between technostress and individual innovation was found to be 0.877, and between individual innovation and perceived OC, it was 0.902. Consequently, the indirect effect of technostress on perceived OC was calculated to be 0.791, demonstrating a robust mediation effect.

$B^{indirect} = a \times b \quad \rightarrow \quad 0/877 \times 0/902 = 0/791$

Referencing Table 5, derived from hypothesis testing results, several significant findings are evident. The first hypothesis, revealing a negative impact of technostress on OC (path coefficient: -0.643, t-value: -1.736), underscores the adverse influence of technostress on employees' commitment levels. The second hypothesis confirms the negative impact of technostress on individual innovation (path coefficient: -0.877, t-value: -25.522), emphasizing the hindrance technostress poses to innovative thinking. Additionally, the third hypothesis establishes a positive correlation between individual innovation and OC (path coefficient: 0.34, t-value: 5.969), indicating that innovation positively impacts organizational commitment. Finally, the fourth hypothesis illustrates the mediating role of individual innovation between technostress and OC, with a path coefficient of 0.644 (Table 6), emphasizing the intermediary function of individual innovation in mitigating the adverse effects of technostress on employees' organizational commitment.

Table (6): The Mediating Effect.

| Dependent variable | Independent variable | β | R | R² | Sig |
|---|---|---|---|---|---|
| Technostress | Perceived OC | 0/940 | 0/940 | 0/883 | 0/000 |
| Technostress | Individual innovation | 0/877 | 0/877 | 0/770 | 0/000 |
| Individual innovation | Perceived OC | 0/902 | 0/902 | 0/814 | 0/000 |
| Technostress, Individual innovation | Perceived OC, | 0/644  0/337 | 0/954 | 0/910 | 0/000 |

Table (7): Results achieved from investigation of hypothesis.

| Hypothesis | Path | Confirmation / rejection |
|---|---|---|
| H 1 | Technostress -> OC | Confirmed |
| H 2 | OC -> individual innovation | Confirmed |
| H 3 | Individual innovation -> OC | Confirmed |
| H 4 | Technostress- individual innovation -> OC | Confirmed |

This study conducted an analysis of two models using Partial Least Squares (PLS) models. The Outer Model, akin to the measurement model, was assessed, while the Inner Model, comparable to the structural model in software such as LISREL, EQS, and AMOS, was scrutinized. The compatibility of the Outer Model was evaluated using Communality, measuring the extent to which the outer model aligns with the structural model. Researchers often consider a communality value greater than 0.5 as acceptable for a statistical subscription (Lee et al., 2008). A communality exceeding 0.5 indicates the model's satisfactory compatibility.

In addition, R2, representing the model's competency in structural description, was examined based on the findings presented in Table 8. R2 signifies the proportion of variance in the dependent variable that is predictable from the independent variables. In this case, R2 serves as an indicator of the model's competence. The results demonstrated that the proposed model exhibits suitable compatibility, validating its adequacy in capturing the underlying relationships and providing a robust structural description of the phenomena under investigation.

Table (8): Model compatibility.

| Variable | Communality | $R^2$ |
|---|---|---|
| Technostress | 0/8704 | |
| Individual innovation | 0/8248 | 0/7687 |
| OC | 0/9446 | 0/9127 |

## 6. Discussion

The present paper has attempted to analyze the impact of technostress on perceived OC by mediating role of individual innovation in an engineering consulting company. Our findings reveal a significant adverse impact of Technostress on Perceived Organizational Commitment (POC) and individual innovation. Simultaneously, individual innovation exerts a positive and significant influence on POC. Additionally, we observed that individual innovation acts as a mediator between technostress and POC, mitigating the negative effects of technostress. These results underline the importance of managerial interventions to alleviate technostress and enhance individual innovation within the organization. Such efforts are crucial for fostering higher levels of organizational commitment among employees.

Regarding our first hypothesis, the factor loading values (11.736) demonstrate a significant negative influence, exceeding 96.1, indicating a correlation between technostress and perceived organizational commitment. This finding aligns with the research of Kumar et al. (2013), which vividly illustrates the negative correlation between technostress, organizational commitment, and job satisfaction. Organizations, especially those immersed in technology, must offer robust technical support. This support should be prompt, available during working hours, and staffed by knowledgeable experts. Limiting unnecessary information entry and emails, sorting information effectively using suitable information systems, and engaging knowledge-based companies for information management are essential steps. Managers should inform employees of approved programs, seek their opinions before adopting new systems, and involve them in aligning the new system with organizational needs and goals. Prior studies, including Harunavamwe and Ward (2022) and Raza et al. (2022) show the negative consequences of technostress for employees, and. They state that technostress leads to diminish employees' wellbeing and performance.

Concerning hypothesis 2, the factor loading values (25.522) signify a significant negative influence, surpassing 96.1, indicating a relationship between technostress and individual innovation. The study findings are in line with the research of Tarafdar et al. (2011) and Ragu-Nathan et al. (2008), validating the negative and significant relationship between technostress and employees' productivity and satisfaction. Therefore, organizations should foster an interactive and open environment, encouraging employees to take risks and gain new experiences. Training and consultation should diminish fear and stress among employees. Managers can enhance participation and reduce technology-induced anxiety by fostering a more collaborative work environment.

Considering hypothesis 3, the factor loading values (5.969) exceed 96.1, demonstrating a significant correlation between individual innovation and perceived organizational commitment. While no prior research explicitly supports this hypothesis, organizations can promote innovation and, consequently, perceived organizational commitment by creating an innovative atmosphere. Encouraging risk-taking and participation, recognizing innovative employees, and fostering a dynamic and competitive atmosphere are vital. Managers must reward and support innovative staff, listen to new ideas, and create a vibrant atmosphere, stimulating individual innovation within the organization.

Regarding hypothesis, the mediation analysis with a path coefficient of 0.644 indicates that individual innovation mediates the relationship between technostress and perceived organizational commitment, reducing the negative impact of technostress. No prior research specifically supports this hypothesis. Therefore, managers should encourage employees to share knowledge related to new technologies, collaborate in problem-solving through group work, adopt up-to-date technologies, innovate new problem-solving approaches, and establish an appropriate reward system. Managers should facilitate the implementation of new and useful ideas, ensuring bureaucratic obstacles do not hinder innovative staff. Open and direct communication is essential for establishing a sense of cooperation and commitment among employees.

### 6.1. Implications for practice and policy

In the face of rapid technological advancements, organizations must strategize to facilitate employee adaptation to new technologies (Nasrollahi et al., 2022), thereby minimizing technostress and maximizing productivity. A multifaceted approach is crucial, involving various aspects of training, support, innovation, learning, and participation to foster a tech-savvy workforce. Moreover, organizations should invest in comprehensive training programs tailored to employees' needs. By focusing on training sessions aimed at individuals requiring new technology skills, organizations can reduce errors and increase efficiency. Encouraging a culture of knowledge sharing among employees can further mitigate technostress. Persuading employees to collaborate in groups, solve problems collectively, and share their expertise helps in reducing the complexity and uncertainty associated with technology.

Providing robust technical support and continuous assistance to employees navigating new technology is pivotal. This assistance can extend to helping employees during crises, ensuring uninterrupted workflow. Specific features such as availability, responsibility, and involving experts can make a significant difference. Supportive measures not only increase employee satisfaction but also reduce the feeling of uncertainty linked to technology. Furthermore, organizations must cultivate an environment that encourages innovation. This involves motivating staff to stay updated with new technologies, create inventive problem-solving approaches, and reward employees who innovate and assist colleagues. Such initiatives not only boost individual innovation but also strengthen organizational commitment. Moreover, promoting a culture of learning and experimentation can significantly reduce technology-induced insecurity. Rewarding employees who seek experiences, learn new skills, and take risks can foster a proactive mindset. Moreover, organizations can enhance employee satisfaction and commitment by involving them actively in technology adoption. This can be achieved by informing employees about new applications, seeking their feedback before implementing new systems, and involving them in tailoring the system to organizational needs. By ensuring the technology aligns with the employees' requirements, organizations increase effectiveness and organizational commitment. Lastly, organizations can mitigate technology-induced overload by implementing measures such as limiting unnecessary data entry and sorting information effectively. Utilizing knowledge-based systems and integrating them into daily operations can streamline information management, reducing technology-induced stress. By adopting these strategies, organizations can create a supportive, adaptive, and innovative tech environment, leading to higher employee satisfaction, increased organizational commitment, and reduced technostress.

### 6.2. Limitations and Suggestions for Future Studies

Limitations of this study are apparent in the confined scope to a single engineering consulting company in Mashhad. Consequently, generalizing these findings to other organizations or similar companies is not viable due to the study's specific context. Control over influential factors like economic and political variables was also limited, potentially affecting research outcomes. Additionally, constraints related to time and financial resources further restricted the study's depth and breadth. Challenges arose due to unresponsive cultures and bureaucratic hurdles in selected companies, hindering the smooth implementation of the research. Limited participant engagement was noted, primarily because of their overwhelming workloads and other commitments. Moreover, some respondents lacked a comprehensive understanding of managerial concepts, compromising the accuracy of their responses.

In terms of future research recommendations, conducting similar studies across diverse organizational settings is crucial to enhance validity and broaden the scope of the findings. These comparative studies would offer valuable insights and allow for more comprehensive analyses. Furthermore, exploring employee satisfaction levels within an engineering consulting company in comparison to similar international entities can provide significant cross-cultural perspectives. Such research could enrich our understanding of organizational commitment and technostress impact in varied work environments, contributing to a more nuanced understanding of these phenomena. The study investigates the impact of technostress on job satisfaction with the mediating influence of individual innovation. It suggests exploring alternative models for each research component and revisiting the research topic to gain deeper insights. Additionally, the study recommends re-examining its findings within a larger statistical population to enhance the robustness and generalizability of the results.

## 7. Conclusion

This study, titled "Investigating the Impact of Technostress on Perceived Organizational Commitment through the Mediating Role of Individual Innovation in an engineering consulting company," has been defined and conducted. Four hypotheses were proposed, and a structural equation model was utilized to either confirm or reject these hypotheses. The methodology and outcomes are detailed in chapters three and four, respectively. The hypotheses were tested on 147 respondents. The path analysis model was employed for analysis, and the results indicated that technostress has a significant negative relationship with organizational commitment. Additionally, technostress has a significant negative association with individual innovation, whereas individual innovation shows a positive and significant relationship with organizational commitment. Furthermore, considering the mediating role of individual innovation between technostress and organizational commitment, it significantly reduces the negative impacts of technostress on organizational commitment.

**Reference**

Al-Jabari, B., & Ghazzawi, I. (2019). Organizational Commitment: A Review of the Conceptual and Empirical Literature and a Research Agenda. *International Leadership Journal, 11*(1).

Allen, N. J., & Meyer, J. P. (1990). The measurement and antecedents of affective, continuance and normative commitment to the organization. *Journal of Occupational Psychology, 63*(1), 1-18.

Argote, L., & Ingram, P. (2000). Knowledge transfer: A basis for competitive advantage in firms. *Organizational Behavior and Human Decision Processes, 82*(1), 150-169.

Arnetz, B. B., & Wiholm, C. (1997). Technological stress: Psychophysiological symptoms in modern offices. *Journal of Psychosomatic Research, 43*(1), 35-42.

Baron, R. M., & Kenny, D. A. (1986). The moderator–mediator variable distinction in social psychological research: Conceptual, strategic, and statistical considerations. *Journal of Personality and Social Psychology, 51*(6), 1173.

Basadur, M. (2004). Leading others to think innovatively together: Creative leadership. *The Leadership Quarterly, 15*(1), 103-121.

Benkhoff, B. (1997). Ignoring commitment is costly: New approaches establish the missing link between commitment and performance. *Human Relations, 50*(6), 701-726.

Boshoff, C., & Mels, G. (1995). A causal model to evaluate the relationships among supervision, role stress, organizational commitment and internal service quality. *European Journal of Marketing, 29*(2), 23-42.

Chen, J., Zhu, Z., & Xie, H. Y. (2004). Measuring intellectual capital: A new model and empirical study. *Journal of Intellectual Capital, 5*(1), 195-212.

Chen, Z. X., & Francesco, A. M. (2003). The relationship between the three components of commitment and employee performance in China. *Journal of Vocational Behavior, 62*(3), 490-510.

Cheng, Y., & Stockdale, M. S. (2003). The validity of the three-component model of organizational commitment in a Chinese context. *Journal of Vocational Behavior, 62*(3), 465-489.

Clugston, M., Howell, J. P., & Dorfman, P. W. (2000). Does cultural socialization predict multiple bases and foci of commitment?. Journal of management, 26(1), 5-30.

Cohen, A. (2007). Commitment before and after: An evaluation and reconceptualization of organizational commitment. *Human Resource Management Review, 17*(3), 336-354.

D'Arcy, J., Gupta, A., Tarafdar, M., & Turel, O. (2014). Reflecting on the "dark side" of information technology use. Communications of the Association for Information Systems, 35(1), 5.

DeLone, W. H., & McLean, E. R. (2003). The DeLone and McLean model of information systems success: A ten-year update. Journal of Management Information Systems, 19(4), 9-30.

Dragano, N., & Lunau, T. (2020). Technostress at work and mental health: Concepts and research results. Current Opinion in Psychiatry, 33(4), 407-413.

Drach-Zahavy, A., Somech, A., Granot, M., & Spitzer, A. (2004). Can we win them all? Benefits and costs of structured and flexible innovation-implementations. Journal of Organizational Behavior, 25(2), 217-234.

Du Plessis, M. (2007). The role of knowledge management in innovation. Journal of knowledge management, 11(4), 20-29.

Elfring, T. (2005). Corporate entrepreneurship and venturing. Springer Science & Business Media. 10.

Foss, N. J. (2007). The emerging knowledge governance approach: Challenges and characteristics. Organization, 14(1), 29-52.

Fuller, J. B., Marler, L. E., & Hester, K. (2006). Promoting felt responsibility for constructive change and proactive behavior: Exploring aspects of an elaborated model of work design. Journal of Organizational Behavior: The International Journal of Industrial, Occupational and Organizational Psychology and Behavior, 27(8), 1089-1120.

Getz, I., & Robinson, A. G. (2003). Innovate or die: Is that a fact? Creativity and Innovation Management, 12(3), 130-136.

Hessari, H., & Busch, P., & Smith, S. (2022a). Supportive leadership and co-worker support for nomophobia reduction: Considering affective commitment and HRM practices. The 33rd Year of Australasian Conference on Information Systems.

Hessari, H., & Nategh, T. (2022b). Smartphone addiction can maximize or minimize job performance? Assessing the role of life invasion and techno exhaustion. Asian Journal of Business Ethics, 11(1), 159–182. https://doi.org/10.1007/s13520-022-00145-2

Hessari, H., & Nategh, T. (2022c). The role of co-worker support for tackling techno stress along with these influences on need for recovery and work motivation. International Journal of Intellectual Property Management, 12(2), 233–259. https://doi.org/10.1504/IJIPM.2022.122301

Janssen, O. (2004). How fairness perceptions make innovative behavior more or less stressful. Journal of organizational behavior, 25(2), 201-215.

Johannessen, J. A., Olaisen, J., & Olsen, B. (2001). Mismanagement of tacit knowledge: The importance of tacit knowledge, the danger of information technology, and what to do about it. International Journal of Information Management, 21(1), 3-20.

Krejcie, R. V., & Morgan, D. W. (1970). Determining sample size for research activities. Educational and psychological measurement, 30(3), 607-610.

Kumar, R., Lal, R., Bansal, Y., & Sharma, S. K. (2013). Technostress in relation to job satisfaction and organizational commitment among IT professionals. International Journal of Scientific and Research Publications, 3(12), 1-3.

Laka-Mathebula, M. R. (2005). Modelling the relationship between organizational commitment, leadership style, human resources management practices and organizational trust (Doctoral dissertation, University of Pretoria).

Lee, J., Park, S. Y., Baek, I., & Lee, C. S. (2008). The impact of the brand management system on brand performance in B–B and B–C environments. Industrial marketing management, 37(7), 848-855.

Mani, K. P., Sritharan, R., & Gayatri, R. (2014). Impact of occupational stress on quality work life among railway station masters of Trichy division. *Bonfring International Journal of Industrial Engineering and Management Science, 4*(4), 165.

Marchiori, D. M., Felix, A. C. S., Popadiuk, S., Mainardes, E. W., & Rodrigues, R. G. (2020). A relationship between technostress, satisfaction at work, organizational commitment and demography: Evidence from the Brazilian public sector. *Revista Gestão & Tecnologia, 20*(4), 176-201.

Meyer, J. P., & Allen, N. J. (1991). A three-component conceptualization of organizational commitment. *Human Resource Management Review, 1*(1), 61-89.

Meyer, J. P., & Allen, N. J. (1997). Commitment in the workplace: Theory, research, and application. *Sage Publications*.

Meyer, J. P., & Herscovitch, L. (2001). Commitment in the workplace: Toward a general model. Human resource management review, 11(3), 299-326.

Meyer, J. P., Paunonen, S. V., Gellatly, I. R., Goffin, R. D., & Jackson, D. N. (1989). Organizational commitment and job performance: It's the nature of the commitment that counts. *Journal of Applied Psychology, 74*(1), 152.

Meyer, J. P., Stanley, D. J., Herscovitch, L., & Topolnytsky, L. (2002). Affective, continuance, and normative commitment to the organization: A meta-analysis of antecedents, correlates, and consequences. *Journal of Vocational Behavior, 61*(1), 20-52.

Miller, W. L., & Morris, L. (2008). Fourth generation R&D: Managing knowledge, technology, and innovation. *John Wiley & Sons*.

Miron, E., Erez, M., & Naveh, E. (2004). Do personal characteristics and cultural values that promote innovation, quality, and efficiency compete or complement each other? *Journal of Organizational Behavior, 25*(2), 175-199.

Mohaghar, A., Ghasemi, R., Toosi, H., & Sheykhizadeh, M. (2022). Evaluating City of Knowledge's Project Management Office Functions Using BWM and Importance-Performance Analysis. *Journal of Decisions & Operations Research, 7*(4).

Mohammadi, A., Ahadi, P., Fozooni, A., Farzadi, A., & Ahadi, K. (2023). Analytical evaluation of big data applications in E-commerce: A mixed method approach. *Decision Science Letters, 12*(2), 457-476.

Moore, J. E. (2000). One road to turnover: An examination of work exhaustion in technology professionals. *MIS Quarterly, 141-168*.

Mowday, R. T., Porter, L. W., & Steers, R. M. (2013). Employee—organization linkages: The psychology of commitment, absenteeism, and turnover. *Academic Press.*

Muthuveloo, R., & Rose, R. C. (2007). Antecedents and outcomes of organizational commitment among Malaysian engineers (Doctoral dissertation, Universiti Putra Malaysia).

Maier, C., Laumer, S., Weinert, C., & Weitzel, T. (2015). The effects of technostress and switching stress on discontinued use of social networking services: A study of Facebook use. *Information Systems Journal, 25*(3), 275-308.

Murray, W. C., & Rostis, A. (2007). Who's running the machine? A theoretical exploration of work stress and burnout of technologically tethered workers. *Journal of Individual Employment Rights, 12*(3), 249-263.

Nasrollahi, M., Ghadikolaei, A. S., Ghasemi, R., Sheykhizadeh, M., & Abdi, M. (2022). Identification and prioritization of connected vehicle technologies for sustainable development in Iran. *Technology in Society, 68*, 101829.

Nawaser, K., Ahmadi, M., Ahmadi, Y., & Dorostkar, M. (2015). Organizational citizenship behavior and bank profitability: Examining relationships in an Iranian bank. *Asian Social Science, 11*(12), 11.

Nawaser, K., Khaksar, S. M. S., Shakhsian, F., & Jahanshahi, A. A. (2011). Motivational and legal barriers of entrepreneurship development. *International Journal of Business and Management, 6*(11), 112.

Nawaser, K., Shahmehr, F. S., Kamel, A., & Vesal, S. M. (2014). Assessing the relationship between strategy and organizational culture in an Iranian manufacturing industry. *Asian Social Science, 10*(21), 175.

Osayawe Ehigie, B., & Clement Akpan, R. (2004). Roles of perceived leadership styles and rewards in the practice of total quality management. *Leadership & Organization Development Journal, 25*(1), 24-40.

Perry, S. J., Hunter, E. M., & Currall, S. C. (2016). Managing the innovators: Organizational and professional commitment among scientists and engineers. *Research Policy, 45*(6), 1247-1262.

Powell, D. M., & Meyer, J. P. (2004). Side-bet theory and the three-component model of organizational commitment. *Journal of Vocational Behavior, 65*(1), 157-177.

Pratama, E. N., Suwarni, E., & Handayani, M. A. (2022). The effect of job satisfaction and organizational commitment on turnover intention with person organization fit as moderator variable. *Aptisi Transactions on Management (ATM), 6*(1), 74-82.

Ragu-Nathan, T. S., Tarafdar, M., Ragu-Nathan, B. S., & Tu, Q. (2008). The consequences of technostress for end users in organizations: Conceptual development and empirical validation. *Information Systems Research, 19*(4), 417-433.

Riedl, R. (2012). On the biology of technostress: Literature review and research agenda. *ACM SIGMIS Database: The DATABASE for Advances in Information Systems, 44*(1), 18-55.

Riggle, R. J., Edmondson, D. R., & Hansen, J. D. (2009). A meta-analysis of the relationship between perceived organizational support and job outcomes: 20 years of research. *Journal of Business Research, 62*(10), 1027-1030.

Rouhani, S., & Mohammadi, A. (2022). A Novel Hybrid Forecasting Approach for Customers Churn in Banking Industry. *Journal of Information & Knowledge Management, 2250089.*

Sasani, F., Mousa, R., Karkehabadi, A., Dehbashi, S., & Mohammadi, A. (2023). TM-vector: A Novel Forecasting Approach for Market stock movement with a Rich Representation of Twitter and Market data. *arXiv preprint arXiv:2304.02094.*

Saunila, M. (2020). Innovation capability in SMEs: A systematic review of the literature. *Journal of Innovation & Knowledge, 5*(4), 260-265.

Senge, P. M. (2006). *The fifth discipline: The art and practice of the learning organization*. Broadway Business.

Sethi, V., Barrier, T., & King, R. C. (1999). An examination of the correlates of burnout in information systems professionals. *Information Resources Management Journal (IRMJ), 12*(3), 5-13.

Siu, O. L. (2003). Job stress and job performance among employees in Hong Kong: The role of Chinese work values and organizational commitment. *International Journal of Psychology, 38*(6), 337-347.

Tarafdar, M., Gupta, A., & Turel, O. (2013). The dark side of information technology use. *Information Systems Journal, 23*(3), 269-275.

Tarafdar, M., Tu, Q., Ragu-Nathan, T. S., & Ragu-Nathan, B. S. (2011). Crossing to the dark side: examining creators, outcomes, and inhibitors of technostress. *Communications of the ACM, 54*(9), 113-120.

Tarafdar, M., Tu, Q., Ragu-Nathan, B. S., & Ragu-Nathan, T. S. (2007). The impact of technostress on role stress and productivity. *Journal of Management Information Systems, 24*(1), 301-328.

Vakola, M., & Nikolaou, I. (2005). Attitudes towards organizational change: what is the role of employees' stress and commitment?. *Employee Relations, 27*(2), 160-174.

Wasti, S. A. (2005). Commitment profiles: Combinations of organizational commitment forms and job outcomes. *Journal of Vocational Behavior, 67*(2), 290-308.

Weil, M. M., & Rosen, L. D. (1997). *Technostress: Coping with technology@ work@ home@ play*. New York: J. Wiley.